\documentstyle[12pt]{article}
\def\bib{\bibitem}
\def\be{\begin{equation}}
\def\ee{\end{equation}}
\def\barr{\begin{array}}
\def\earr{\end{array}}
\def\lsim{\:\raisebox{-0.5ex}{$\stackrel{\textstyle<}{\sim}$}\:}
\def\gsim{\:\raisebox{-0.5ex}{$\stackrel{\textstyle>}{\sim}$}\:}
\def\GeV{\:{\rm GeV}}
\def\lbar{\overline{\lambda}}
\def\vti{v^{(t)}_i}
\def\vtj{v^{(t)}_j}
\def\vei{v^{(e)}_i}
\def\vej{v^{(e)}_j}
\def\ati{a^{(t)}_i}
\def\atj{a^{(t)}_j}
\def\aei{a^{(e)}_i}
\def\aej{a^{(e)}_j}
\def\sW{\sin\theta_W}
\def\cW{\cos\theta_W}
\def\ib#1,#2,#3{           {\it ibid.\/ }{\bf #1} (19#2) #3}
\def\ap#1,#2,#3{           {\it Ann. Phys. (NY)\/ }{\bf #1} (19#2) #3}
\def\ijmp#1,#2,#3{         {\it Int. J. Mod. Phys.\/ } {\bf A#1} (19#2) #3}
\def\mpl#1,#2,#3 {          {\it Mod. Phys. Lett.\/ } {\bf A#1} (19#2) #3}
\def\np#1,#2,#3{           {\it Nucl. Phys.\/ }{\bf B#1} (19#2) #3}
\def\npps#1,#2,#3{         {\it Nucl. Phys. B (Proc. Suppl.)\/ }{\bf B#1}
                             (19#2) #3}
\def\plb#1,#2,#3{           {\it Phys. Lett.\/ }{\bf B#1} (19#2) #3}
\def\pr#1,#2,#3{           {\it Phys. Rev.\/ }{\bf D#1} (19#2) #3}
\def\prep#1,#2,#3{         {\it Phys. Rep.\/ }{\bf #1} (19#2) #3}
\def\prl#1,#2,#3{          {\it Phys. Rev. Lett.\/ }{\bf #1} (19#2) #3}
\def\pro#1,#2,#3{          {\it Prog. Theor. Phys.\/ }{\bf #1} (19#2) #3}
\def\rmp#1,#2,#3{          {\it Rev. Mod. Phys.\/ }{\bf #1} (19#2) #3}
\def\sp#1,#2,#3{           {\it Sov. Phys.-Usp.\/ }{\bf #1} (19#2) #3}
\def\zp#1,#2,#3{           {\it Zeit. f\"ur Physik\/ }{\bf #1} (19#2) #3}
\def\etal{ {\it et al.} }
\def\ie{ {\it i.e.} }
\def\eg{ {\it e.g.} }

\begin{document}
\thispagestyle{empty}
\setcounter{page}{0}
\renewcommand{\thefootnote}{\fnsymbol{footnote}}

\begin{flushright}
{\bf hep-ph/9408250} \\[4ex]

MPI-PhT/94-50\\
August 1994
\end{flushright}

\vspace{5ex}
\begin{center}

{\Large \bf Leptoquark Search at $e^+ e^-$ Colliders}\\

\bigskip
\bigskip
{\sc
   Debajyoti Choudhury\footnote{debchou@iws186.mppmu.mpg.de,
debchou@dmumpiwh.bitnet}
   }

\bigskip
{\it Max-Planck-Institut f\"ur Physik, Werner-Heisenberg-Institut,
F\"ohringer Ring 6, \\
80805 M\"unchen, Germany.} \\

\bigskip
\bigskip
{\bf Abstract}
\end{center}

\begin{quotation}
We investigate the possibility of detecting a scalar leptoquark, coupling
to the electron and the top, at a linear collider. For coupling strength
equalling the weak coupling constant, the present mass bounds are of the
order of 300 GeV. We demonstrate that at the NLC, one could detect such
particles if their mass were less than a few TeV's.
\end{quotation}

\vspace{8ex}

\newpage
\setcounter{footnote}{0}
\renewcommand{\thefootnote}{\arabic{footnote}}

In the otherwise glorious success of the Standard Model (SM),
there remains the
``aesthetic drawback'' of the arbitrary assignment of quark and
lepton fields. Theories that venture beyond the SM do offer hints of a
pattern \cite{guts,composite} albeit at the cost of new fields that mediate
new interactions between the quarks and leptons. Leptoquarks are but a kind.
Transforming as the fundamental representation of the $SU(3)_c$ gauge group,
these can be either (pseudo-)scalars or (axial-)vectors.

Some phenomenogical constraints (at least for leptoqaurks coupling to the
first two generations) are obvious. The stability of the proton
dictates that if the mass ($m$) and the coupling ($f$)
be such that $m/f \lsim 10^{16}
\GeV$, then a  leptoquark may not have a diquark coupling\cite{proton}.
Bounds\cite{fcnc} from flavour
changing neutral current processes ($m/f \gsim 10^5 \GeV$) are
evaded only if we assume that these couple to only one flavour
of leptons and quarks each\cite{leurer}. Furthermore, the requirement that
the leptonic decay modes of the pseudoscalar mesons be not enhanced stipulates
that either  $m/f \gsim 10^5 \GeV$ or the leptoqaurk couples
 chirally\cite{chiral}.
Bounds from atomic parity violation are somewhat weaker\cite{leurer}.

Inspite of such constraints, there has been a recent surge of interest in the
subject\footnote{For a representative list of recent work, please
see ref.\protect\cite{gg}}.
 The reasons are twofold. On the one hand, there do exist symmetries
and models wherein the above constraints are comfortably evaded while
allowing a rich phenomenology; on the other, recent experiments have begun
to provide us with direct probes\cite{direct}. For coupling strengths similar
to the weak gauge coupling, these searches give lower bounds of the order
of a few hundred GeV.

There could exist though one class of leptoquarks that would easily evade
the aforementioned constraints without any further assumptions.
We focus here on a scalar particle ($S$) that
couples the electron only with the  quarks of the
third generation. In the SM, the ordinary Yukawa
coupling for the top quark is almost the same as the weak gauge coupling. If
this be the case here too, one could look forward to interesting new features.
In a recent work, Bhattacharyya \etal\cite{gg}, have investigated such a
scenario. They conclude that the strongest bounds can be inferred from
the leptonic partial widths of the $Z$, and for the above choice of
coupling strengths, are of the order of 300 GeV.

In this letter we point out that at the next generation of linear
colliders (such as CLIC,JLC,TESLA,VLEPP {\it etc.})
a significant improvement can be made. While direct production obviously
limits us to the kinematic bound, one could probe larger masses by
considering interactions involving virtual leptoquarks. With the discovery
of the top quark and the determination of its mass\cite{top},
a simple experiment suggests itself.

The additional piece in the Lagrangian that is of relevance to us can be
parametrized in the form
\be
  {\cal L}_Y = g \: S \:\bar{t} (h_L P_L + h_R P_R) e
\ee
where $g$ is the weak coupling constant and $h_{L,R}$ are dimensionless
constants.
%  From the discussion above we know that only one of the two could
%  be non-zero.

Let us now concentrate on the interaction
$e^-(p_1,\lambda) e^+(p_2,\lbar) \longrightarrow t(p_3) \bar{t}(p_4)$. Here
$\lambda,\lbar$ denote the polarization of the electron and the positron
respectively. Within the SM, the interaction proceeds {\it via} two
tree--level $s$-channel diagrams, the amplitude for which can be expressed as
\be
\displaystyle
{\cal M}_{SM} =
 \sum_i   \frac{g^2}{s - m_i^2}
   \bar{u}(p_3)\: \gamma_\mu \left( \vti + \ati \gamma_5 \right) \:v(p_4) \;\:
   \bar{v}(p_2,\lbar)\: \gamma_\mu \left( \vei + \aei \gamma_5 \right) \:
             u(p_1,\lambda)
\ee
where the sum runs over the photon and the $Z$ and
\be
\barr{rclcrcl}
v^{(e)}_\gamma & = & - \sW &\quad& a^{(e)}_\gamma & = & 0 \\[2ex]
v^{(e)}_Z & = & \displaystyle \frac{-1 + 4 \sin^2\theta_W}{4 \cW}
               &\quad& a^{(e)}_Z & = & \displaystyle \frac{1}{4 \cW}
                                 \\[2ex]
v^{(t)}_\gamma & = & \displaystyle \frac{2\sW}{3}
                        &\quad& a^{(t)}_\gamma & = & 0 \\[2ex]
v^{(t)}_Z & = & \displaystyle \frac{1 - 8 \sin^2\theta_W/3}{4 \cW}
               &\quad& a^{(t)}_Z & = & \displaystyle \frac{-1}{4 \cW}\\
\earr
\ee
With the introduction of the leptoquark, we have an additional $t$--channel
diagram :
\be
\displaystyle
{\cal M}_{LQ} =
 \frac{g^2}{t - m_S^2}
   \bar{u}(p_3)\: \left( h_L P_L + h_R P_R\right) \:u(p_1,\lambda) \;\:
   \bar{v}(p_2,\lbar)\: \left( h_L^\ast P_R + h_R^\ast P_L \right) \:
             v(p_4).
\ee

The very fact that the additional contribution is a  $t$--channel one as
opposed to the  $s$--channel ``background'', points to the possibility of a
significant modification in the angular distribution. That this is indeed so
is borne out by a glance at the differential cross section. For reasons of
compactness, we first define:
\be
\barr{rcl}
\displaystyle A_{ij} & = & \displaystyle
      (1 + \lambda \lbar) \left\{ \vei \vej + \aei \aej \right\}
            - (\lambda + \lbar) \left\{ \vei \aej + \aei \vej \right\} \\[2ex]
\displaystyle B_{ij} & = & \displaystyle
      (1 + \lambda \lbar) \left\{ \vei \aej + \aei \aej \right\}
            - (\lambda + \lbar) \left\{ \vei \vej + \aei \vej \right\} \\[2ex]
\displaystyle C_{ij} & = & \displaystyle \vti \vtj + \ati \atj \\[2ex]
\displaystyle D_{ij} & = & \displaystyle \vti \atj + \ati \vtj \\[2ex]
% \displaystyle E_{ij} & = & \displaystyle \vti \vtj - \ati \atj \\[2ex]
\displaystyle Q_{i} & = & \displaystyle
          |h_L|^2 (1 + \lambda) (1 + \lbar) \left(\vei - \aei \right)
                                            \left(\vti + \ati \right)\\[1.5ex]
        & + & \displaystyle
           |h_R|^2 (1 - \lambda) (1 - \lbar) \left(\vei + \aei \right)
                                            \left(\vti - \ati \right)
   \\[2ex]
\displaystyle R_{i} & = & \displaystyle
          |h_L|^2 (1 + \lambda) (1 + \lbar) \left(\vei - \aei \right)
                                           \left(\vti - \ati \right)\\[1.5ex]
        & + & \displaystyle
            |h_R|^2 (1 - \lambda) (1 - \lbar) \left(\vei + \aei \right)
                                            \left(\vti + \ati \right)
   \\[2ex]
\displaystyle F & = & \displaystyle
         \frac{1}{4}
            \left[ |h_L|^2 (1 + \lambda) + |h_R|^2 (1 - \lambda) \right]
              \:  \left[ |h_L|^2 (1 + \lbar) + |h_R|^2 (1 - \lbar) \right] ~.

\earr
\ee
Finally we have, in terms of the Mandelstam variables,
\be
\barr{rcl}
\displaystyle \frac{ {\rm d} \sigma} { {\rm d} t }& = & \displaystyle
              \frac{3 \pi \alpha^2}{s^2\: \sin^4 \theta_W}
         \left[ 2 \sum_{ij}
           \frac{A_{ij} C_{ij} ( u^2 + t^2 - 2 m_t^4)
                   + B_{ij} D_{ij} s (t - u) + 4  \vti \vtj A_{ij} m_t^2 s }
                                             {(s - m_i^2) (s - m_j^2)}
         \right. \\[2ex]
&  & \displaystyle \hspace*{6em}
       + \left.
             \sum_{i} \frac{Q_i (t - m_t^2)^2 + R_i m_t^2 s}
                           {(t - m_S^2) (s - m_i^2)}
             + F \left( \frac{t-m_t^2}{t - m_S^2} \right)^2
        \right]~.
\earr
   \label{diff c.s.}
\ee

A straightforward integration of the above expression would obviously give
us an indication of the observability at a collider of a given beam energy.
A more sensitive test could be to compare the observed forward--backward
asymmetry with the SM expectation. Much more can be achieved  though by
comparing the angular
distributions. To do this, one would divide
 the angular width of the experiment
into
bins and compare the observed number of events $n_j$ in each with the SM
prediction $n_j^{SM}$. A $\chi^2$ test can be devised as
\be \displaystyle
    \chi^2 = \sum_{j=1}^{ {\rm bins} }
      \left( \frac{ n_j^{SM} - n_j}{\Delta n_j^{SM} } \right)^2
\label{chi2}
\ee
The number of events obviously is obtained by integrating
 eqn.(\ref{diff c.s.}) over the part of the
phase space corresponding to the particular
bin and is given by
\be
    n_i = \sigma_i \epsilon L
\ee
where $\epsilon$ is the detector efficiency and $L$ is the machine luminosity.
The error in eqn.(\ref{chi2})
 is a combination of the statistical and systematic ones
{\it viz.}
\be
\Delta n = \sqrt{ (\sqrt{n})^2 + (\delta_{\rm syst} n)^2 }.
  \label{error}
\ee

To make our results quantitative, we choose to investigate for
 C.M. energies of 500 GeV and 1000 GeV. For the
integrated luminosity we assume the oft-quoted figure\cite{LC92} of
 10 fb$^{-1}$. For experimental simplicity, we delimit ourselves to be
at least $20^\circ$ away from the beam pipe. The efficiency ($\epsilon$) for
the top-reconstruction is taken to be 15\% and we make a conservative
estimate of $\delta_{\rm syst} = 0.05$. Dividing the angular region into
10 equal-sized
bins\footnote{We find that the sensitivity of the results to the binning
is rather weak for bin cardinality between 6 and 20.},
we then perform the $\chi^2$ test as in
eqn(\ref{chi2}). To avoid spurious results,
we reject a bin from the  analysis if either
$(i)$    the difference between the SM expectation
	and the measured number of events
	is less than one
or
$(ii)$   the SM expectation is less than one event
	while the measured number is less than three.

For reasons of clarity, we do not attempt to present our results as a function
of all three of the parameters $m_S, h_L, h_R$.
We rather choose to present these as bounds in the two--parameter
space of  mass and one of the couplings, keeping the other
zero. The interpretation is straightforward. Any combination of the
two
parameters {\em above} the curves (\ie away from the origin) can be ruled
out at 95\% C.L.\footnote{If the value of one of the parameters were known,
then a 98.6\% C.L. bound on the other is given by the corresponding projection
on the axis.} Since we are dealing with one--sided bounds,
this corresponds to $\chi^2 > 4.61$ in eqn.(\ref{chi2}).
Figure 1 shows the bounds for the ``left--handed'' leptoquark
for both $\sqrt{s} = 500 \GeV$ and 1000 GeV.

If $|h_L| = 1$, we would then be able to detect (with unpolarized beams)
the corresponding particle
at NLC500 if its mass were less than 2.4 TeV, and at NLC1000 if $m_S < 3.75$
TeV. This should be compared with the current bounds of $m_S \gsim 300 \GeV$
as obtained in ref.\cite{gg}.

In Fig. 1, we also indicate the bounds that would be accessible if beam
polarization were achieved, a distinct possibility at the NLC. Since
polarizing positrons is a relatively difficult proposition, we have chosen
$\lbar = 0$ in eqn.(\ref{diff c.s.}). For simplicity, we present curves
only for $\lambda = 0.5$ and $\lambda = 1.0$ (\ie 50\% and 100\%
left--polarized electron beams). A glance at the figure shows that this
would improve the detection capability by a significant factor.

A corresponding analysis for the right--chirally coupled field can be made
as easily and we present the results thereof in Fig. 2. In this case, we
would obviously do better with a right--polarized electron beam and hence
we present our results for $\lambda = 0,\:-0.5,\:-1.0$.
A comparison of Figs. 1 and 2 show that, for unpolarized beams,
the obtainable limits for the
right--chiral field are weaker compared to those for the left--chiral
one. This can easily be traced to the significant difference between the
left--handed and the right--handed couplings of the $t$--quark to $Z$ leading
to a different level of ``background''. This difference obviously narrows
down as the extent of  beam polarization increases. We summarize our results
in Table 1.

Until now, we have restricted ourselves to the case where only one of
$h_{L,R}$ is non-zero. This was motivated by the consideration that if the
leptoquark is relatively light, then its coupling to the first two generations
must necessarily be chiral. For our case, this constraint is obviously  not
so strict.
In Figs. 3 and
4, we present the 95\% C.L. bound in the $h_L$--$h_R$ plane ( for
 representative values of $m_S = 3$ TeV and $m_S = 4$ TeV )
that can be achieved at the NLC500 and NLC1000 respectively.
The curves for 100\% beam polarization have been dispensed with as these
would obviously be parallel to the coordinate axes.

To conclude, we point out that the present constraints on a scalar
leptoquark that couples the electron and the top are indirect and relatively
 weak.  However at the next linear collider, significant improvement can be
made  in constraining the properies of such particles. Indeed, for a coupling
strength close to the weak coupling constant, such a collider would be able
to detect such particles such particles of masses upto about four times the
center of mass energy.

 {\bf Acknowledgements} I wish to thank G. Bhattacharyya for useful
suggestions and a careful reading of the manuscript. Thanks are due to
F.~Cuypers for useful comments and for sharing his numerical codes.

\newpage
\begin{table}[htb]
\vspace*{4ex}
$$
\begin{array}{||c||c|c|c||c|c|c||}
\hline
 & \multicolumn{6}{c||}{\rm Limit~ of~ Observability~ (TeV)} \\
\cline{2-7}
{\rm Energy}& \multicolumn{3}{c||}{ |h_L| = 1, \quad |h_R| = 0 }
             & \multicolumn{3}{c||}{ |h_R| = 1, \quad |h_L| = 0 } \\
\cline{2-7}
& \lambda = 0 & \lambda = 0.5 & \lambda = 1.0
       & \lambda = 0 & \lambda = -0.5 & \lambda = -1.0 \\
\hline \hline
\sqrt{s} = 500~{\rm GeV} & 2.39 & 2.75 & 3.02 &1.95 & 2.50 & 3.10 \\
\hline
\sqrt{s} = 1000~{\rm GeV} & 3.75 & 4.31 & 4.73 & 2.98 & 3.76 &4.66  \\
\hline
\end{array}
$$

\vspace{2ex}

\caption{Limits of observability (98.6\% C.L.) for scalar leptoquarks
coupling chirally
to the
electron and the top-quark with a strength equalling the weak coupling
constant. $\lambda$ is the electron polarization, the positron beam being
unpolarized.}
\end{table}

%\newpage
\begin{figure}[htb]
\vskip 8in\relax\noindent\hskip -1in\relax{\includegraphics{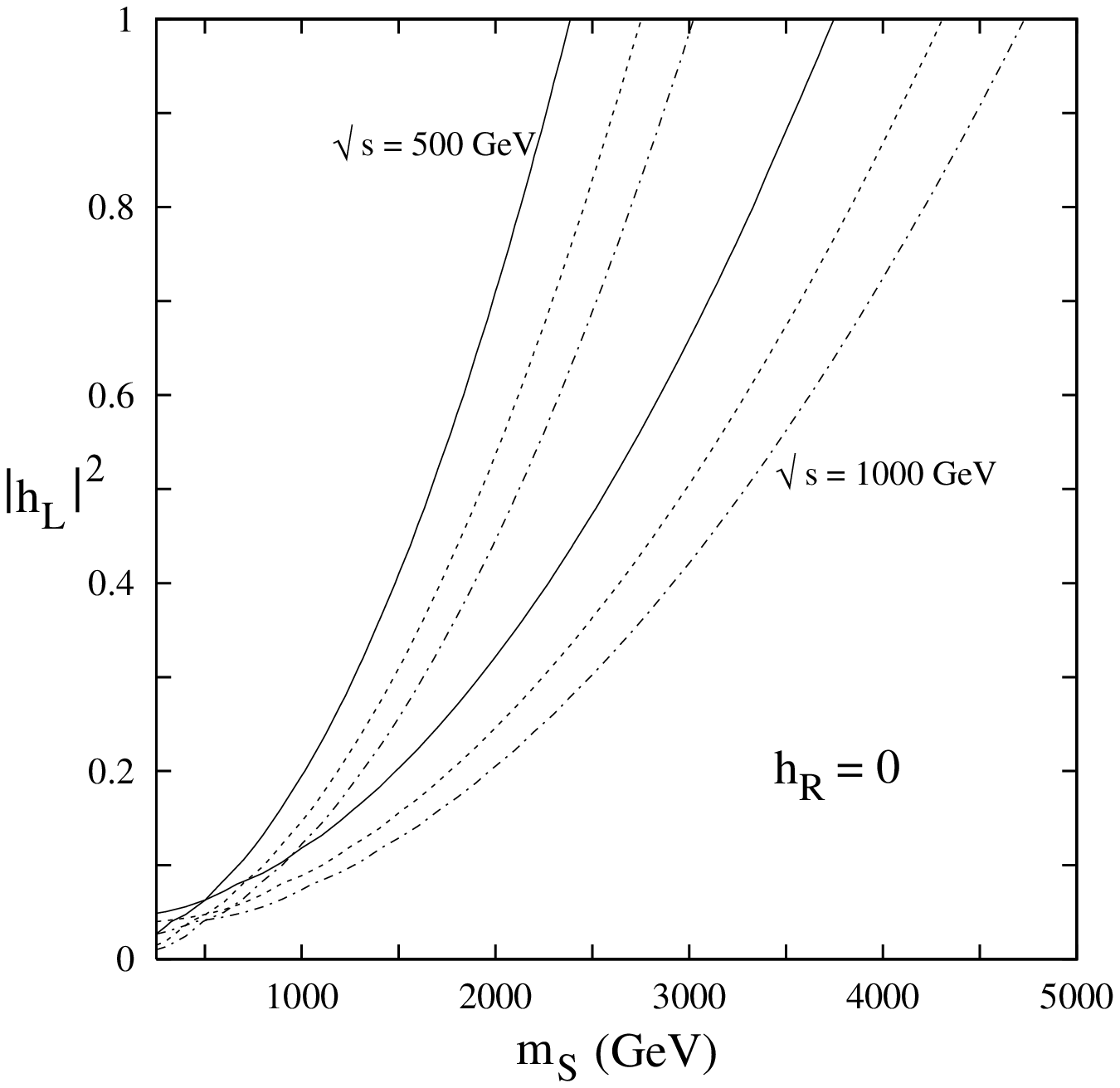}}

\vspace{-20ex}
\caption{Contours of detectability in the $m_S$--$h_L$ plane for $h_R = 0$
(left set : NLC500, right set : NLC1000).
The parameter space {\em above} the curves can be ruled out at 95\% C.L. The
solid, dashed and dot-dashed curves are for electron polarization 0, $+ 0.5$
 and $+ 1.0$ respectively (the positron is unpolarized). }
       \label{fig1}
\end{figure}

\begin{figure}[htb]
\vskip 8in\relax\noindent\hskip -1in\relax{\includegraphics{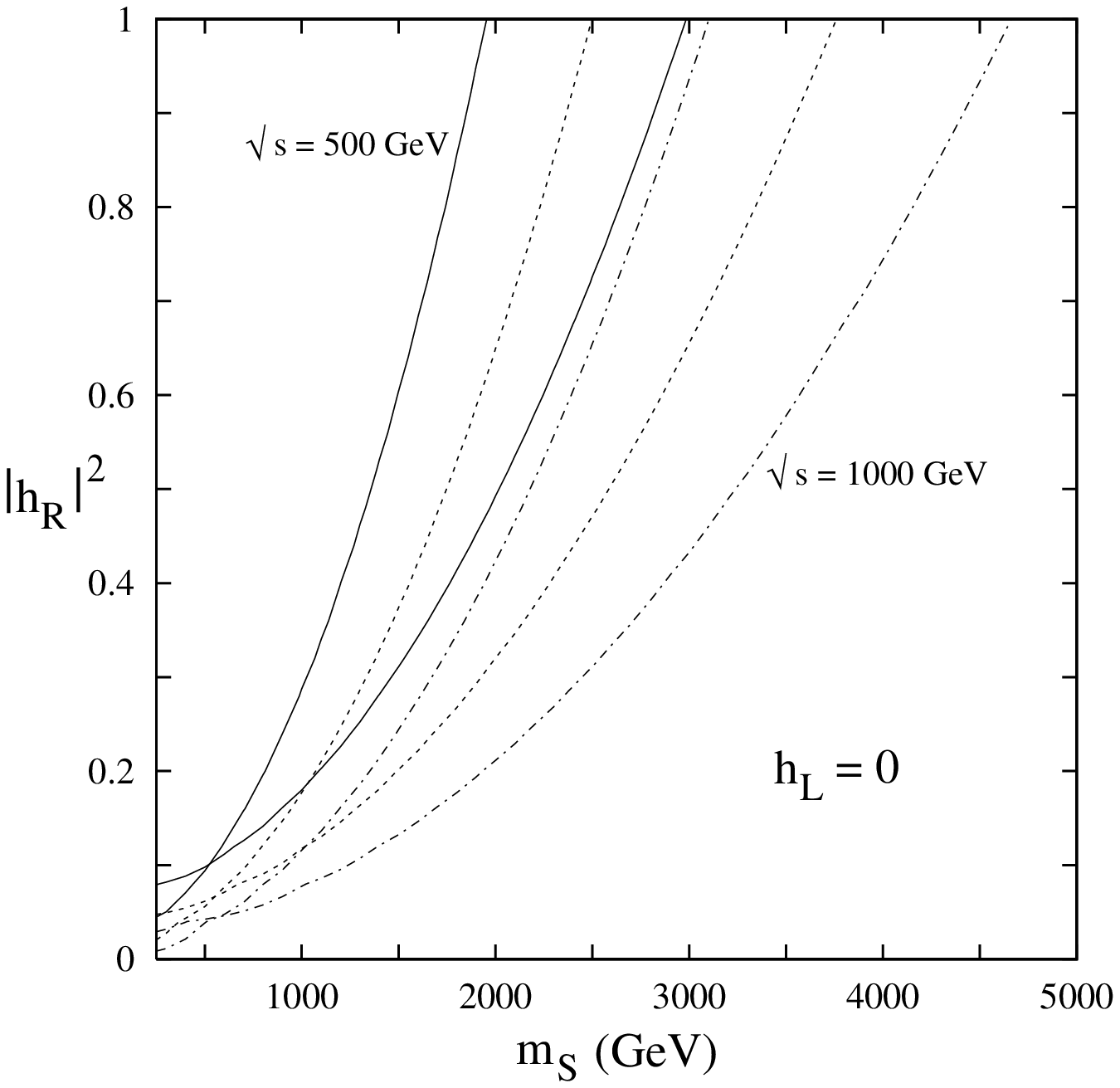}}

\vspace{-20ex}
\caption{
   Contours of detectability in the $m_S$--$h_R$ plane for $h_L = 0$
(left set : NLC500, right set : NLC1000).
The parameter space {\em above} the curves can be ruled out at 95\% C.L. The
solid, dashed and dot-dashed curves are for electron polarization 0, $- 0.5$
 and $- 1.0$ respectively (the positron is unpolarized).
}
       \label{fig2}
\end{figure}

\begin{figure}[htb]
\vskip 8in\relax\noindent\hskip -1in\relax{\includegraphics{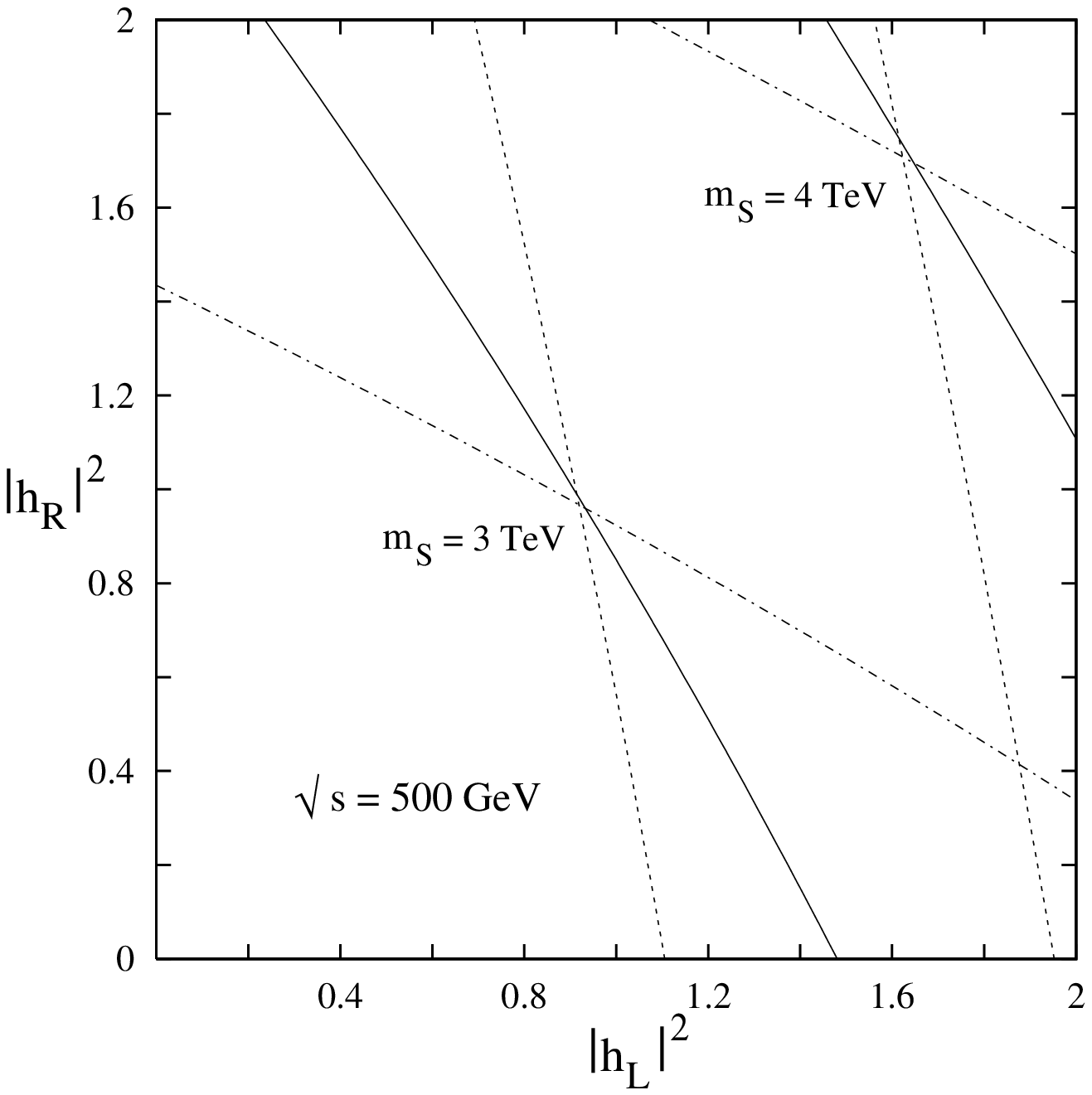}}

\vspace{-20ex}
\caption{
   Contours of detectability in the $h_L$--$h_R$ plane at NLC500.
The parameter space {\em above} the curves can be ruled out at 95\% C.L. The
solid, dashed and dot-dashed curves are for electron polarization 0, $+ 0.5$
 and $- 0.5$ respectively (the positron is unpolarized).
}
       \label{fig3}
\end{figure}

\begin{figure}[htb]
\vskip 8in\relax\noindent\hskip -1in\relax{\includegraphics{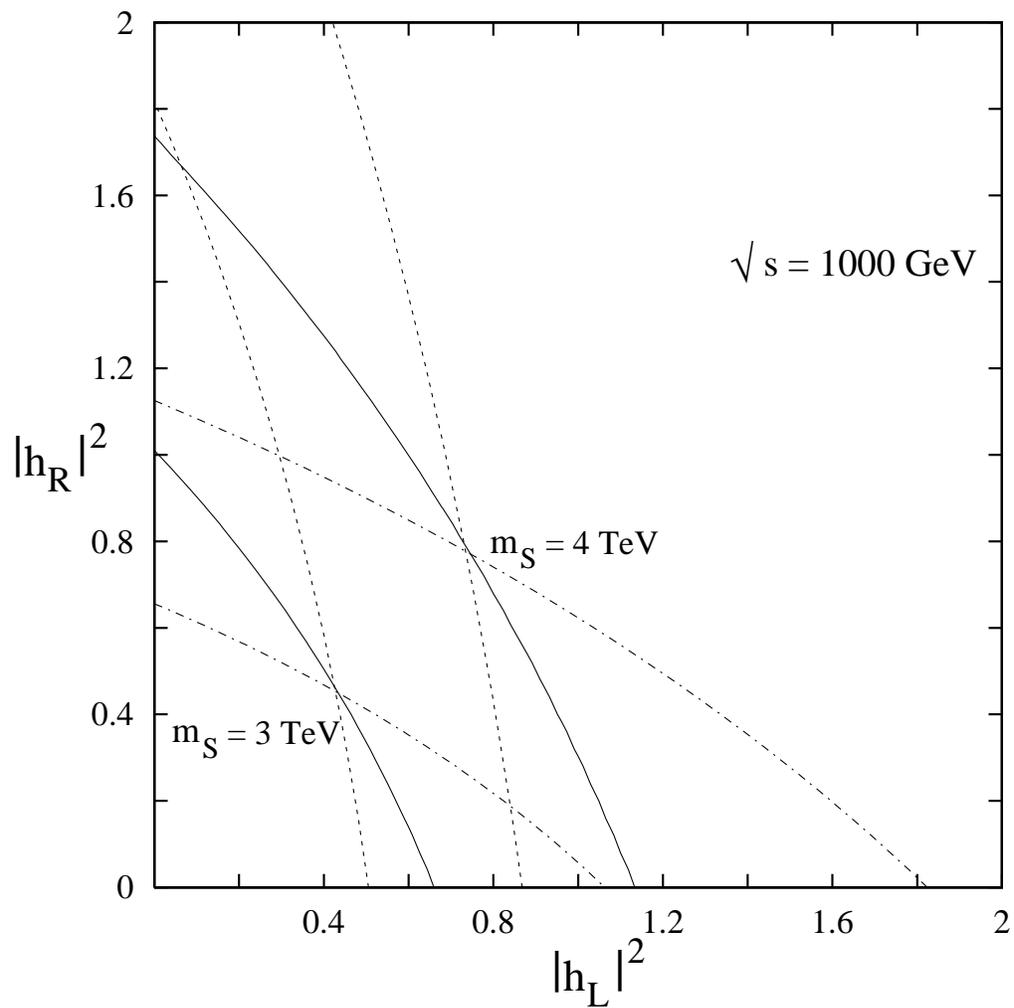}}

\vspace{-20ex}
\caption{   As in Fig. 3, but for NLC1000.
}
       \label{fig4}
\end{figure}

\end{document}